# LIGHT CONTROLLED PHOTON TUNNELING

Igor I. Smolyaninov [1], Anatoliy V. Zayats [2], and Christopher C. Davis [1]

[1] *Department of Electrical and Computer Engineering*

*University of Maryland, College Park, MD 20742*

*Phone: 301-405-3255, Fax: 301-314-9281, e-mail: smoly@eng.umd.edu*

[2] *Department of Pure and Applied Physics, Queen's University of Belfast,*

*Belfast, BT7 1NN, United Kingdom*

**Abstract**

Recent measurements of photon tunneling through individual subwavelength pinholes in a gold film covered with a layer of polydiacetylene (Phys. Rev. Letters **88**, 187402 (2002)) provided strong indication of "photon blockade" effect similar to Coulomb blockade phenomenon observed in single-electron tunneling experiments. Here we report first observation of photon tunneling been blocked (gated) by light at a different wavelength. This observation suggests possibility of building new class of photon tunneling gating devices for all-optical signal processing.

Very recently [1] we have reported strong experimental and theoretical evidence of "photon blockade" effect in light tunneling through individual subwavelength pinholes in a gold film covered with a layer of polydiacetylene, which may occur even at a level of single photon tunneling. This effect is somewhat similar to Coulomb blockade phenomenon observed in single-electron tunneling experiments. Coulomb blockade leads to single-electron tunneling in tunnel junctions with an extremely small capacitance, where the charging energy $e^2/2C$ of the capacitance $C$ is much larger than the thermal energy $k_BT$ and the quantum fluctuation energy $h/RC$, where $R$ is the resistance of the tunnel junction[2,3]. Classical realization of light tunneling is based on a glass surface illuminated in the total internal reflection geometry, e.g., using a prism or a semicylinder. In this case, all incident light is reflected and only an evanescent field (exponentially decaying from the surface) exists over a smooth surface. If a tapered glass fiber is placed sufficiently close to the glass-air interface, the evanescent field is transformed into propagating waves in the fiber. Thus, optical tunneling through an air gap (which can be considered as a tunnel barrier) occurs. When a nanometer-scale nonlinear optical object which exhibits well defined localized electromagnetic modes is placed in the tunneling gap, photon tunneling may be facilitated if the frequency of tunneling photons coincides with the frequency of the localized electromagnetic mode. On the other hand, excitation of such modes may cause local changes in the dielectric constant of the nonlinear object:

$$\varepsilon = \varepsilon_0 + 4\pi\chi^{(3)}|E_L|^2 \qquad (1)$$

where $E_L$ is the local field, and $\varepsilon_0$ and $\chi^{(3)}$ are the linear dielectric constant and the third order nonlinear susceptibility of the object, respectively. As a consequence of dielectric constant change, the localized modes may experience a noticeable frequency change, so that the tunneling photons will not be in resonance any more. Thus, photon tunneling will be blocked in a manner similar to the Coulomb blockade effect for electrons.

Our measurements of light tunneling through individual subwavelength pinholes in a thick gold film covered with a layer of poly-3-butoxy-carbonyl-methyl-urethane (3BCMU) polydiacetylene provided strong evidence of photon blockade effect. A thick (~0.5 µm) gold film had been thermally deposited on the surface of a glass prism. Microscopic images of the film revealed usual granular structure of the film with the size of individual grains in the 10-100 nm range. Only a few sparsely separated pinholes were visible in the film under illumination with 488 nm and 632 nm light at an angle larger than the angle of total internal reflection for the glass-air interface. The pinholes selected for the measurements had very low optical transmission of a few hundred photons per second The size $a$ of these pinholes may be estimated as a few nanometers using the Bethe's expression

$$S \sim a^2(a/\lambda)^4 \qquad (2)$$

for the cross section of a subwavelength aperture [4], and the data for the pinhole transmission. A drop of 3BCMU polydiacetylene solution in chloroform [5] was deposited onto the gold film surface. After solvent evaporation a thick film of polydiacetylene was left on the surface. 3BCMU and 4BCMU polydiacetylene materials hold the current record for the largest fast nonresonant optical $\chi^{(3)}$ nonlinearity [6]. In combination with a very strong local field enhancement expected for surface plasmons localized between nanometer scale gold particles (the pinholes are presumed to be located at the grain boundaries), our theoretical estimates [1] give reasonable chances of photon blockade observation even at the level of single photon tunneling. While the transmission dependencies on the illuminated light intensity measured for most pinholes were linear, in accordance with our estimates optical transmission of some small pinholes exhibited saturation and even highly nonlinear staircase-like behavior. The observed nonlinearity suggests that at least at some frequencies $\omega_1$ and $\omega_2$ of the incident light, tunneling of $\omega_1$ light through such small pinholes may be blocked (gated) by $\omega_2$ light incident on the same pinhole. In this Letter we report first experimental observation of such light controlled photon tunneling. This observation suggests possibility of building new class of photon tunneling gating devices for all-optical signal processing.

Our experimental setup is shown in Fig.1. A bent optical fiber tip similar to the bent tips often used in near-field optical microscopes [7] was used to collect light transmitted through the individual pinholes in the sample described above. The fiber tip was positioned above individual pinholes in contact with the thick polymer coating of the sample using a far-field optical microscope and the shear-force distance control system commonly used in near-field optical setups [8]. Since pinholes were typically separated by at least 10 μm distances, properties of individual pinholes were studied in the absence of any background scattered light. Under the far-field microscope control the same ~30x30 μm$^2$ area of the sample containing a few pinholes has been illuminated with 488 nm light of a CW argon ion laser and 632 nm light of a CW He-Ne laser. The far-field microscopic images of two pinholes homogeneously illuminated with 632 nm light (a) and 488 nm light (b) of comparable intensity are shown in Fig.2. These images indicate difference in the optical properties of the pinholes probably related to their geometry. While both pinholes show about the same transmission at 632 nm, one pinhole in Fig.2(b) is almost invisible at 488 nm. In our previous paper [1] the pinhole geometry was modeled as a narrow gap $d$ between two small metal spheres (representing nanometer-size grains of the polycrystalline film). The nonlinear material is assumed to fill the gap. The localized plasmon modes spectrum of such a system can be calculated analytically in the limit $d<<R$ since it formally resembles particle motion in the Coulomb field [9], and is determined by the following dispersion relation

$$Re(\varepsilon / \varepsilon(\omega)) = -(m+1/2)(d/2R)^{1/2}, \ m = 0, 1, 2 ..., \quad (3)$$

where $\varepsilon$ and $\varepsilon(\omega)$ are the dielectric constants of nonlinear material and metal, respectively, and $R$ is the radius of the sphere. The quantum number $m$ corresponds to different localized surface plasmon modes in the gap. This well-known spectrum suggests considerable differences in the optical properties of the pinholes in the visible range, so our experimental observations in Fig.2 are not surprising.

We have studied how transmission at 632 nm of the pinhole visible at both wavelengths in Fig.2 is affected by simultaneous illumination with 488 nm light. The data of pinhole transmission at 632 nm as a function of illuminated light intensity measured with and without auxiliary 488 nm illumination are shown in Fig.3(a). Measured without 488 nm illumination, the transmission of the pinhole exhibits weak nonlinearity consistent with the much larger estimated pinhole size (of the order of 30 nm) compared to a few-nanometer-size pinholes studied in ref.[1]. The transmission measured with 488 nm illumination has been consistently lower (in agreement with the photon blockade hypothesis), although large statistical noise does not allow to make such a conclusion with absolute certainty. In order to observe light controlled tunneling directly, we modulated incident 488 nm light with a chopper at a frequency of 1.3 kHz. Light transmitted through the pinhole was collected with the fiber and sent to a photomultiplier (PMT) through an optical filter, which completely cuts of the blue light. Modulation of the pinhole transmission at 632 nm induced by modulation of 488 nm light incident on the same pinhole was observed using a lock-in amplifier (Fig.1), which used the chopper sync signal as a reference. Thus, small variations of 632 nm light tunneled through the pinhole induced by 488 nm light have been measured directly. Results of this experiment are shown in Fig.3(b). The shutter of the He-Ne laser was closed and opened two times during the experiment. In order to prove that this effect is related to the nonlinear properties of the pinhole covered with 3BCMU, we have conducted similar measurements on an uncoated pinhole with similar optical properties located on the same sample (Fig.3(c)). In this case no tunneling light modulation has been detected.

In conclusion, we have reported first observation of 632 nm photon tunneling through a subwavelength pinhole been blocked (gated) by light at a different (488 nm) wavelength. This observation suggests possibility of building new class of photon tunneling gating devices for all-optical signal processing. In this application low optical throughput of individual pinholes may be compensated by large number of pinholes per device area.

# References

[1] I.I. Smolyaninov, A.V. Zayats, A. Gungor, and C.C. Davis, *Phys.Rev.Lett.* **88**, 187402 (2002).

[2] E. Ben-Jacob and Y. Gefen, *Phys.Lett. A* **108**, 289 (1985).

[3] D.V. Averin and K.K. Likharev, *J.Low Temp.Phys*. **62**, 345 (1986).

[4] H. Bethe, *Phys.Rev*. **66**, 163 (1944).

[5] We are grateful to A. Drury and W. Blau, Polymer Research Centre, Trinity College Dublin, for kindly providing the 3BCMU polymer used in the experiment.

[6] A. K. Yang, W. Kim, A. Jain, J. Kumar, S. Tripathy, *Optics Comm*. **164**, 203 (1999).

[7] A. Lewis, K. Lieberman, N. BenAmi, G. Fish, E. Khachatryan, U. BenAmi, and S. Shalom, *Ultramicroscopy*, **61**, 215 (1995).

[8] E. Betzig, P. L. Finn, J. S. Weiner, *Appl.Phys.Lett*. **60**, 2484 (1992).

[9] V. M. Agranovich, V. E. Kravtsov, T. A. Leskova, *Sol.State Comm*. **47**, 925 (1983).

**Figure Captions**

Figure 1. Schematic view of our experimental setup.

Figure 2. 30x50 $\mu m^2$ images of the same area of the sample illuminated with 632 nm light (a) and 488 nm light (b) of comparable intensity. These images indicate different optical properties of the two pinholes visible in the field of view.

Figure 3. (a) Transmission of the left (Fig.2) pinhole at 632 nm as a function of illuminated light intensity measured with (stars) and without (squares) auxiliary 488 nm illumination. This pinhole exhibits weak nonlinearity consistent with its size. (b) Modulation of the left pinhole transmission at 632 nm light induced by modulation of 488 nm light incident on the same pinhole. Modulation of 488 nm light was provided by the chopper (Fig.1) at a frequency of 1.3 kHz. The shutter of the He-Ne laser was closed and opened two times during the experiment. (c) Similar experiment performed on a pinhole which has not been covered with 3BCMU polydiacetylene.

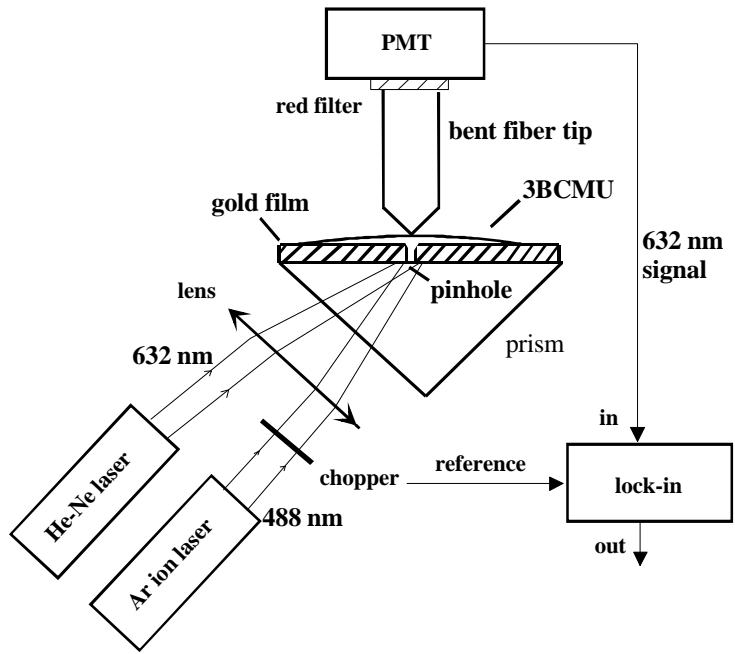

Figure 1.

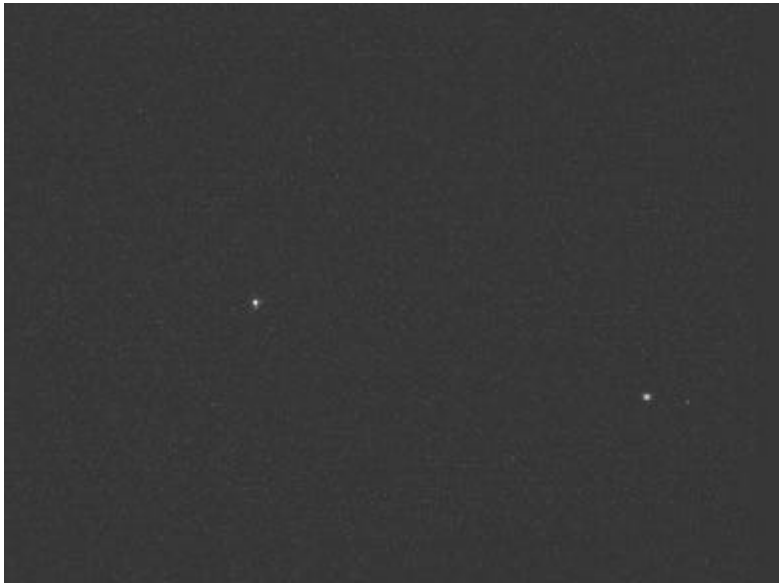

**(a)**

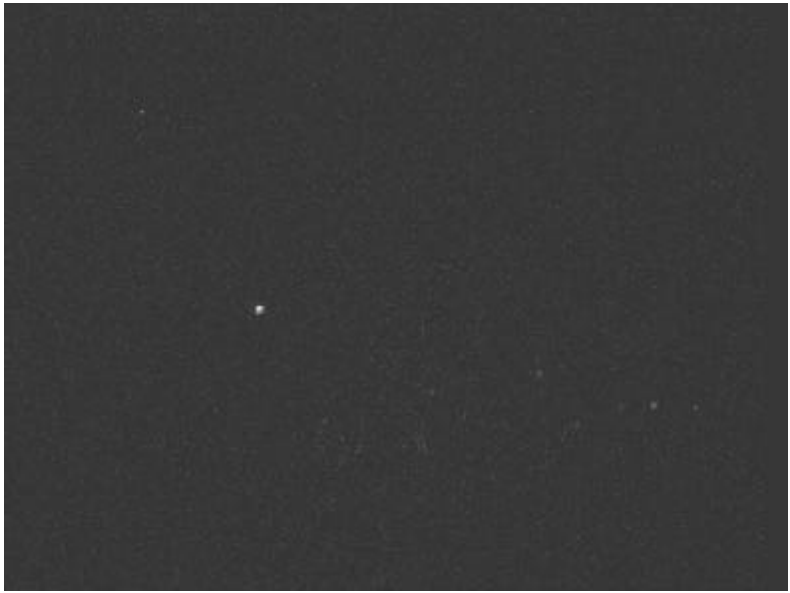

**(b)**

Figure 2

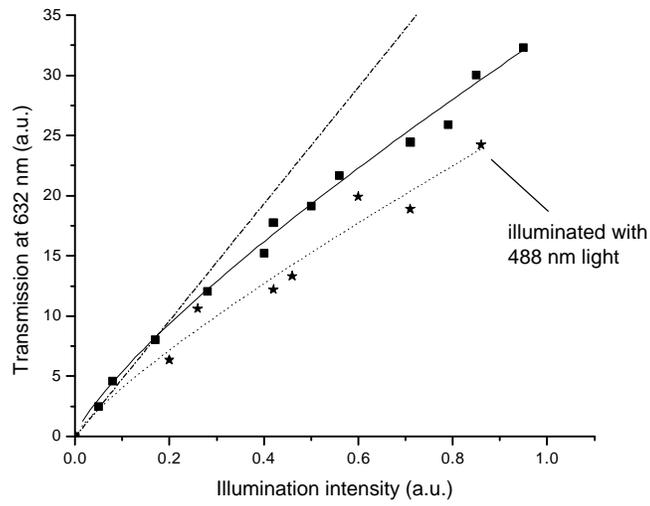

**(a)**

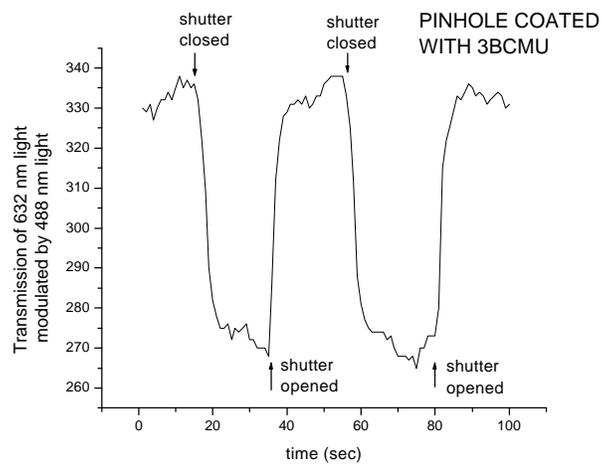

**(b)**

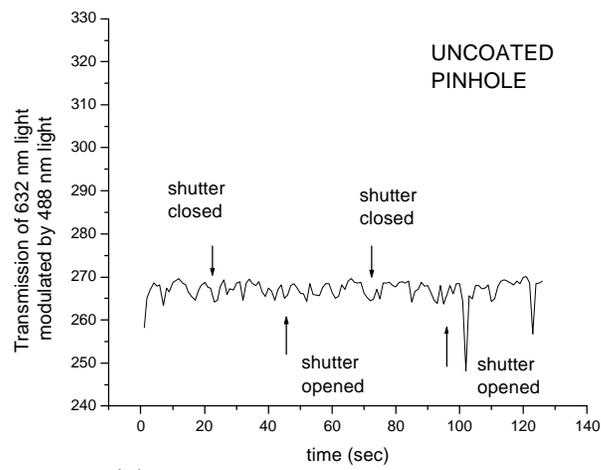

**(c)**

Figure 3.